\documentclass[prl,amsmath,amssymb,twocolumn, superscriptaddress]{revtex4-1}
\usepackage{graphicx}

\usepackage{color}

\newcommand{\gdot}{\dot{\gamma}}

\begin{document}
\preprint{APS/123-QED}

\begin{abstract}
Shear thickening is a widespread phenomenon in suspension flow that, despite sustained study, is still the subject of much debate. The longstanding view that shear thickening is due to hydrodynamic clusters has been challenged by recent theory and simulations suggesting that contact forces dominate, not only in discontinuous, but also in continuous shear thickening. 
Here, we settle this dispute using shear reversal experiments on micron-sized silica and latex colloidal particles to measure directly the hydrodynamic and contact force contributions to shear thickening. We find that contact forces dominate even continuous shear thickening. Computer simulations show that these forces most likely arise from frictional interactions.
\end{abstract}

\title{Hydrodynamic and contact contributions to shear thickening in colloidal suspensions}
\author{Neil Y. C. Lin}
\affiliation{Department of Physics, Cornell University, Ithaca, NY 14853, USA}
\author{Ben M. Guy}
\affiliation{SUPA, School of Physics and Astronomy, University of Edinburgh, Edinburgh EH9 3FD, United Kingdom}%
\author{Michiel Hermes}
\affiliation{SUPA, School of Physics and Astronomy, University of Edinburgh, Edinburgh EH9 3FD, United Kingdom}%
\author{Chris Ness}
\affiliation{School of Engineering, University of Edinburgh, Edinburgh EH9 3JL, United Kingdom}%
\author{Jin Sun}
\affiliation{School of Engineering, University of Edinburgh, Edinburgh EH9 3JL, United Kingdom}%
\author{Wilson C. K. Poon}
\affiliation{SUPA, School of Physics and Astronomy, University of Edinburgh, Edinburgh EH9 3FD, United Kingdom}%
\author {Itai Cohen}
\affiliation{Department of Physics, Cornell University, Ithaca, NY 14853, USA}
\email{yl834@cornell.edu}
\date{\today}
\maketitle

Shear thickening, the increase of viscosity with shear rate, is ubiquitous in complex fluids \cite{Cheng2011, Wagner2009, mewis2012colloidal, Brown2014}. In particular, it plays a controlling role in the flow of concentrated suspensions of micron-sized particles~\cite{Guy2015}. Such suspensions occur widely in applications, from ceramics and bullet-proof armor, through cement and even chocolate. Shear thickening has also become a mainstay of popular science, in the form of running on a pool of corn starch solution. 

Despite sustained study, the mechanism of suspension shear thickening is still disputed. A long-standing view is that thickening, especially if it is continuous with shear rate, is predominantly driven by hydrodynamic interactions \cite{Bossis1989, Brady1997, Wagner2009, mewis2012colloidal}. In some theories, the increase in viscosity arises directly from enhanced dissipation in the narrow lubrication films between particles \cite{Brady1997, melrose1996continuous}; in others, it is caused by the formation of transient particle clusters (``hydroclusters") \cite{raghavan1997shear, Bossis1989, Wagner2009, mewis2012colloidal, Brady1985}. In support, experiments have identified putative hydroclusters~\cite{Cheng2011, Kalman2009, maranzano2002flow} and found an increase in hydrodynamic stresses during thickening~\cite{Bender1996,Gurnon2015,xu2014microstructure}. 

Recently, contact forces have been shown to mediate a discontinuous jump in viscosity with shear rate in dense suspensions of non-Brownian particles at volume fractions greater than random loose packing $\phi \gtrsim 0.58$~\cite{Brown2014, fernandez2013microscopic, fall2015macroscopic, fall2008shear, brown2009dynamic, brown2010generality}. More controversially, simulations \cite{Seto2013, Mari2014} and theories \cite{Bashkirtseva2009, Wyart2014} suggest that direct frictional contact can also lead to continuous shear thickening in moderately dense colloids with $\phi \lesssim 0.58$. This proposal has not yet gained wide acceptance: conceptually, it seems harder to find a role for such forces without the formation of system-spanning contact networks~\cite{Brown2014}. 

Experiments that can dissect the relative contributions of hydrodynamics and contact stresses can settle this issue definitively. Here, we demonstrate that shear reversal techniques, pioneered by Gadala-Maria \emph{et al}.~\cite{Gadala-Maria1980,Kolli2002}, can distinguish the relative contributions of these stresses in non-inertial shear thickening systems. The basic idea is simple but powerful: immediately upon shear reversal, instantaneous contact stresses vanish, but hydrodynamic stresses remain because of Stokes flow reversibility. Thus, monitoring time-dependent stresses after reversal will reveal the relative roles of these two interactions in thickening.  

\begin{figure}[t]
\centering
\includegraphics[width=0.3\textwidth]{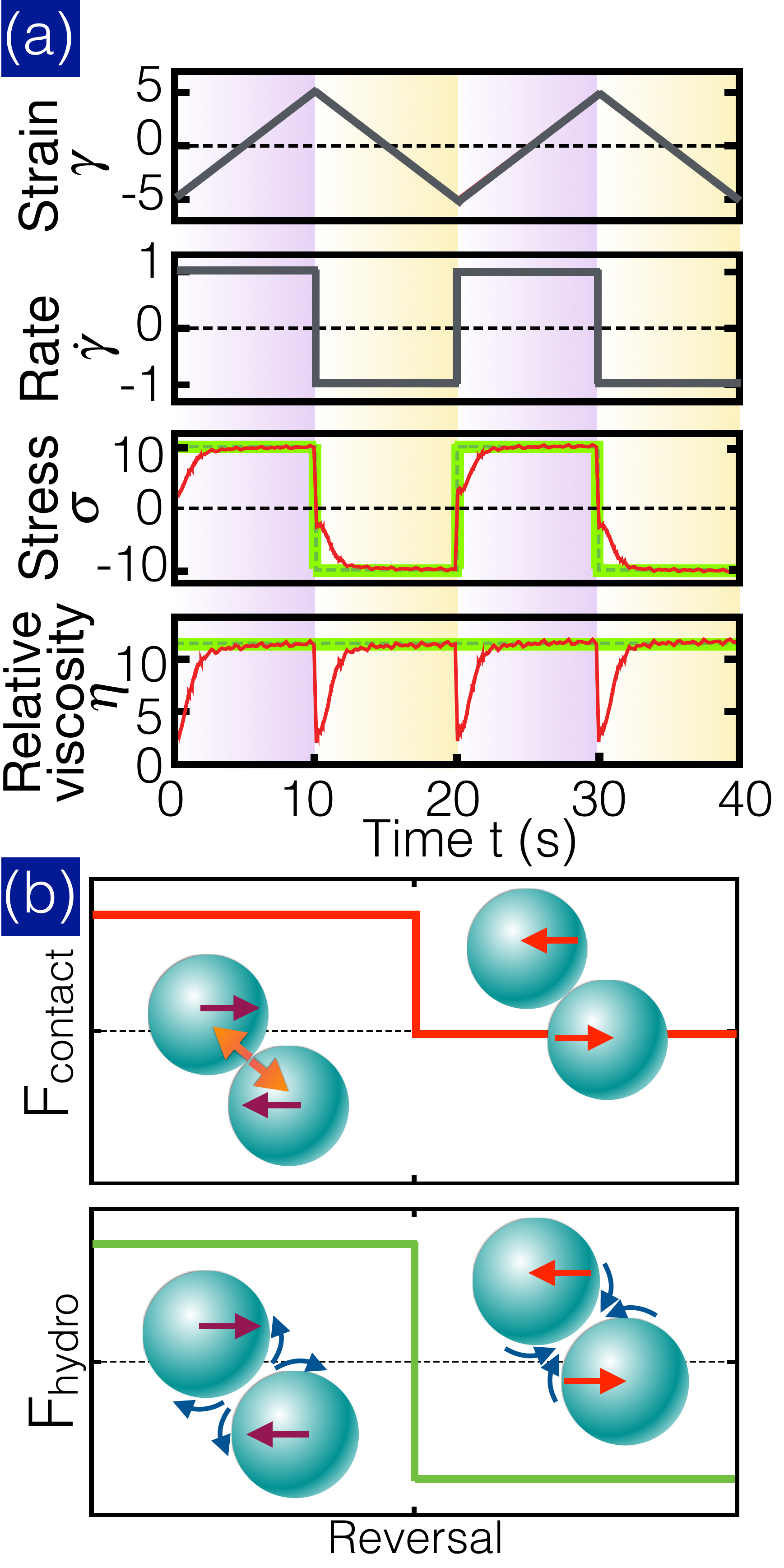}
\caption{
(a) Schematic of our shear reversal protocol. The applied strain
$\gamma$ is a triangle wave with peak-to-peak amplitude of 10. The applied
strain rate $\gdot(t)$ is a square wave whose amplitude we vary in our experiments. The instantaneous stress
response, $\sigma$, of a viscous fluid and a contact-dominated fluid are shown in the
green and red, respectively. (The same color scheme applies throughout all parts of this figure.)
The instantaneous relative viscosity $\eta(t)=\sigma(t)/(\gdot(t)\eta_0)$. 
(b) Schematic illustrating the difference between hydrodynamic forces, which
stay the same in magnitude immediately upon reversal, and contact forces, which drop to zero immediately upon reversal. 
\label{fig:fig1_schematics}}
\end{figure}

We do so in two canonical colloidal systems: charge-stabilized silica (Seikisui Chemical) with diameter $2.0$ $\mu$m suspended in a
mixture of glycerol and water (viscosity $\eta_0 = 0.98$ Pa.s at 20$^{\circ}$C) at volume fraction $\phi = 0.49$, and polymethylmethacrylate (PMMA) particles with diameter $1.4$ $\mu$m sterically stabilized with poly-dimethyl-diphenyl siloxane (PDV-2335, Gelest) with chain length $\approx 50$ nm \cite{Kogan2008} suspended in PDV-2331 (Gelest, viscosity $\eta_0 =1.78$ Pa.s at 20$^{\circ}$C ) at $\phi = 0.51$. 

Shear reversal measurements were performed in an ARES strain-controlled
rheometer (Rheometric Scientific) with roughened cone-plate geometry (25 mm, 0.1 rad) modified with a DAQ directly connected to the
analogue output of the stress and strain sensors \cite{dullaert2005thixotropy}.
Directly measuring the output of these sensors bypasses
signal processing that hinders the instantaneous measurement of the system's
short-time response. 

In our protocol, figure~\ref{fig:fig1_schematics}(a), a positive
shear rate $\gdot$ is imposed until the accumulated strain $\gamma$ reaches 10, after which the
shear is reversed and a negative shear rate is imposed to accumulate the same amount of strain. Just before shear reversal, the suspension is in a steady state. Immediately upon reversal, the structure is unchanged, so that hydrodynamic forces will remain
identical in magnitude but reversed in direction. In contrast, contact forces in hard-sphere systems will drop to zero immediately after reversal, figure~\ref{fig:fig1_schematics}(b). This qualitative difference between these two forces allows us to disentangle their separate contributions by measuring the transient stress upon reversal. Figure~\ref{fig:fig1_schematics}(a) illustrates schematically the instantaneous relative viscosity 
$\eta(t)=\sigma(t)/(\dot{\gamma}(t)\eta_0)$
as a function of time (or, equivalently, accumulated strain) of a purely
Newtonian fluid (green line) and the response where contact forces
contribute to the total stress (red line).
 (See Supplemental Information for calibration.)

\begin{figure}[t]
\centering
\includegraphics[width=0.48\textwidth]{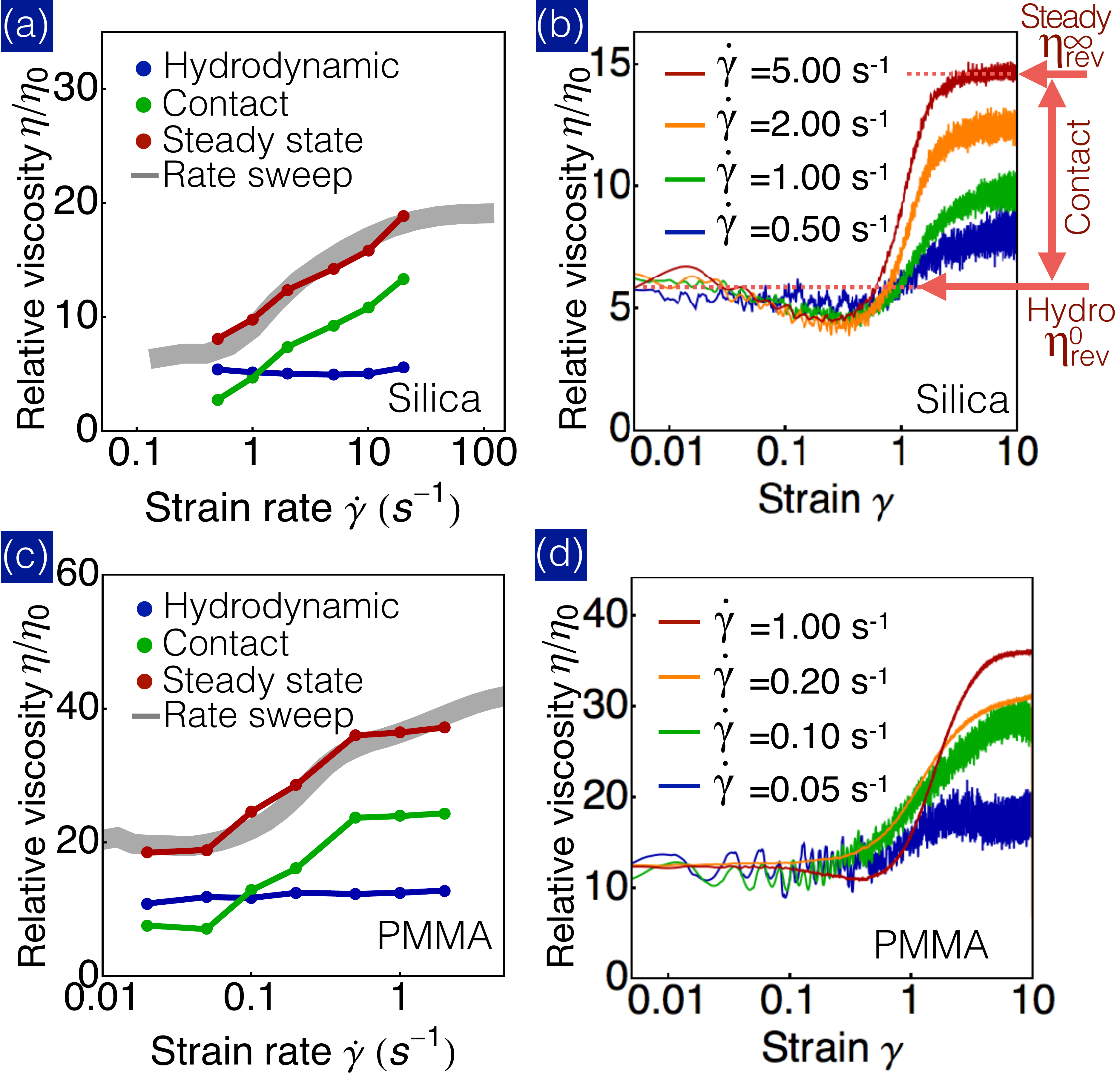}
 \caption{Relative viscosity, $\eta$, \emph{vs} strain rate, $\gdot$, for the (a) silica and (b) PMMA suspensions. Shown are rate sweep measurements (thick grey line), steady state relative viscosities from flow reversal (red) decomposed into contributions from hydrodynamic (blue) and contact (green) interactions. (b) and (d): Instantaneous relative viscosity after shear reversal, $\eta_{rev}(\gamma)=\sigma(\gamma)/(\gdot \eta_0)$, \emph{vs} strain, $\gamma$, at different applied $\gdot$. The hydrodynamic component of the relative viscosity, $\eta_{rev}^0$, is taken to be the average of the relative viscosity over the range $0.01 < \gamma < 0.2$. The contact component is taken to be the difference between the steady state value at $\gamma=10$, $\eta_{rev}^{\infty}$, and $\eta_{rev}^0$.}
\label{fig:fig2_exp_reversal}
\end{figure}

\begin{figure*}[t]
\centering
\includegraphics[width=0.75 \textwidth]{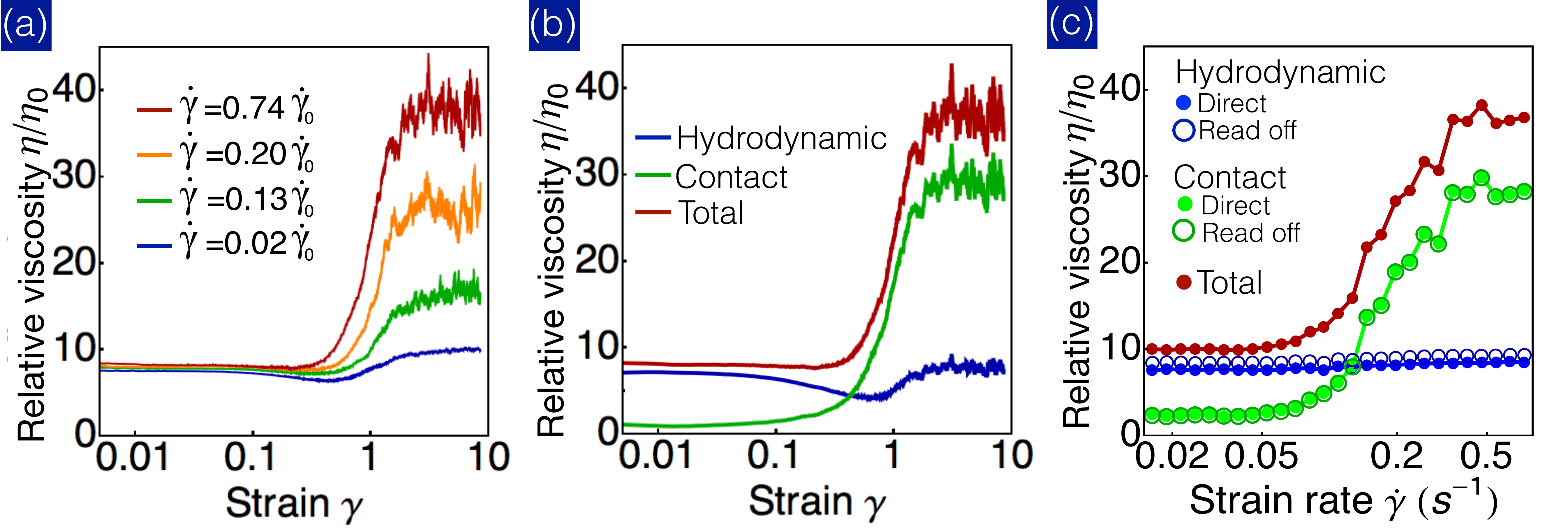}
\caption{Critical load model simulations of particles with 
hydrodynamic and frictional interactions. (a) Relative
viscosity $\eta(\gamma)$ versus strain $\gamma$ at four different shear
rates (given in the legend, see text for the definition of $\dot\gamma_0$); compare experimental results in figure~\ref{fig:fig2_exp_reversal}(b) and (d). (b) Directly-calculated hydrodynamic and contact contributions to the total viscosity as functions of the accumulated strain after reversal for the case of $\dot{\gamma}=0.74 \gdot_0$. (c) The total viscosity as a function of shear rate, and its decomposition into hydrodynamic and contact contributions done in two ways: directly calculated from simulations, and `read off' the data shown in part (a) in the same way as the experimental data is analyzed into these contributions, for which see figure~\ref{fig:fig2_exp_reversal}(b).}
\label{fig:sim}
\end{figure*}

A rate-sweep measurement of the charge-stabilized silica suspension gives its flow curve, $\eta(\dot\gamma)$,
figure~\ref{fig:fig2_exp_reversal}(a) (thick grey line), which shows shear thickening
at $\dot{\gamma} \gtrsim 0.3$~s$^{-1}$. Next, we monitor the viscosity after shear reversal, $\eta_{\rm rev}$, as a function of accumulated strain after reversal, $\gamma$, at four representative shear rates $\dot{\gamma} = 0.50, 1.00, 2.00$ and $5.00$~s$^{-1}$, in the shear thickening regime, figure~\ref{fig:fig2_exp_reversal}(b). Immediately before reversal, $\eta$ has a value  (not shown) equal to that measured in the steady state ($\gamma = 10$). The viscosity immediately after reversal drops to a value, $\eta_{\rm rev}^0$, that remains constant, to within experimental error, until $\gamma \gtrsim 0.1$. We take $\eta_{\rm rev}^0$ to be the hydrodynamic contribution to the total steady-state viscosity. From $\gamma \approx 0.3$, $\eta_{\rm rev}$ rises, reaching a steady-state value, $\eta_{\rm rev}^\infty$, that is the same as the steady-state viscosity before reversal. We take $\eta_{\rm rev}^\infty - \eta_{\rm rev}^0$ to be the contact contribution. 

The hydrodynamic and contact contributions to the total viscosity so obtained are plotted as a function of $\dot\gamma$ in figure~\ref{fig:fig2_exp_reversal}(a). Strikingly, while the contact contribution increases with $\dot\gamma$, the hydrodynamic contribution remains constant as the
suspension shear thickens. This demonstrates the essential role played by contact forces in the continuous shear thickening of this silica suspension.

Contact forces depend sensitively on the nature of particle surfaces. We therefore repeated our experiments using a PMMA suspension. In contrast to our silica particles, which are charge stabilized, these PMMA particles are stabilized by (neutral) surface `hairs'. Our $\phi = 0.51$ PMMA suspension shear thickens more readily, at $\dot\gamma \approx 0.01$ s$^{-1}$, figure~\ref{fig:fig2_exp_reversal}(c). Again using shear-reversal to measure $\eta_{\rm rev}$, figure~\ref{fig:fig2_exp_reversal}(d), and analyzing the data in the same way to obtain the hydrodynamic and contact contributions, figure~\ref{fig:fig2_exp_reversal}(c), we find essentially the same pattern of behavior already uncovered for the $\phi = 0.49$ silica suspension. Thus, contact forces also dominate the continuous shear thickening in this sterically-stabilized system. 

To validate this interpretation, and to probe the nature of these contact forces, we conducted computer
simulations of repulsive spheres (radius $a$) in which the short-range lubrication and repulsive contact forces were mimicked using a previously-established critical load model  \cite{Seto2013}, implemented in a classical discrete element method code (details of which are identical to those in \cite{Ness2015}; also, see Supplemental Information). In this model, frictional interactions appear beyond a critical normal force between particles, $F^{CL}$, which also sets a shear rate scale, $\dot{\gamma}_0 = F^{CL}/ (6\pi \eta_0 a^2)$. We explored shear reversal at $\phi=0.51$. 

The observed evolution of the viscosity with accumulated strain after reversal, figure~\ref{fig:sim}(a), is qualitatively similar to experiments. In the simulations, however, we can follow the hydrodynamic and contact contributions as a function of accumulated reversed strain by direct evaluation; figure~\ref{fig:sim}(b) shows the results for $\dot{\gamma}=0.74\dot\gamma_0$. Consistent with the interpretation we have offered for our experimental data, the contact contribution drops essentially to zero immediately after reversal, and only increases back to its steady-state magnitude after an accumulated stain of order unity. By contrast, the hydrodynamic contribution remains more or less constant. 

We cannot access directly the hydrodynamic and contact stresses in our experiments. Instead, we have to `read off' these contributions to the reversed viscosity from the data as $\eta_{\rm rev}^0$ and $\eta_{\rm rev}^\infty - \eta_{\rm rev}^0$, respectively, figure~\ref{fig:fig2_exp_reversal}(b). We validate this procedure using simulations. First, we `read off' the hydrodynamic and contact contributions to the reversed viscosity data, figure~\ref{fig:sim}(a), in exactly the same way as in experiments, giving results, figure~\ref{fig:sim}(c), analogous to the experimental data shown in figure~\ref{fig:fig2_exp_reversal}(a) and (c). Secondly, we evaluated these contributions directly from the raw simulation data, and overlaid the results in the same graph, figure~\ref{fig:sim}(c). The near identity of the two sets of results validates our identification of $\eta_{\rm rev}^0$ and $\eta_{\rm rev}^\infty - \eta_{\rm rev}^0$ with the hydrodynamic and contact contributions, respectively. Thus, simulations confirm the key role played by contact forces in continuous shear thickening.

Traditionally, shear thickening is supposed to occur beyond a critical onset strain rate \cite{Barnes1989}. Recent theories, simulations and experiments \cite{Seto2013,Mari2014,melrose1996continuous, Wyart2014,Guy2015} suggest that, instead, the phenomenon appears at a critical onset shear stress. Indeed, this assumption has been built into the critical load model used in our simulations. To explore the role of stress in the onset of shear thickening, we performed rate sweep measurements in our two systems at different temperatures, $T$, to change the solvent viscosity, $\eta_0(T)$, over an order of magnitude.

\begin{figure}[t]
\centering
\includegraphics[width=0.48\textwidth]{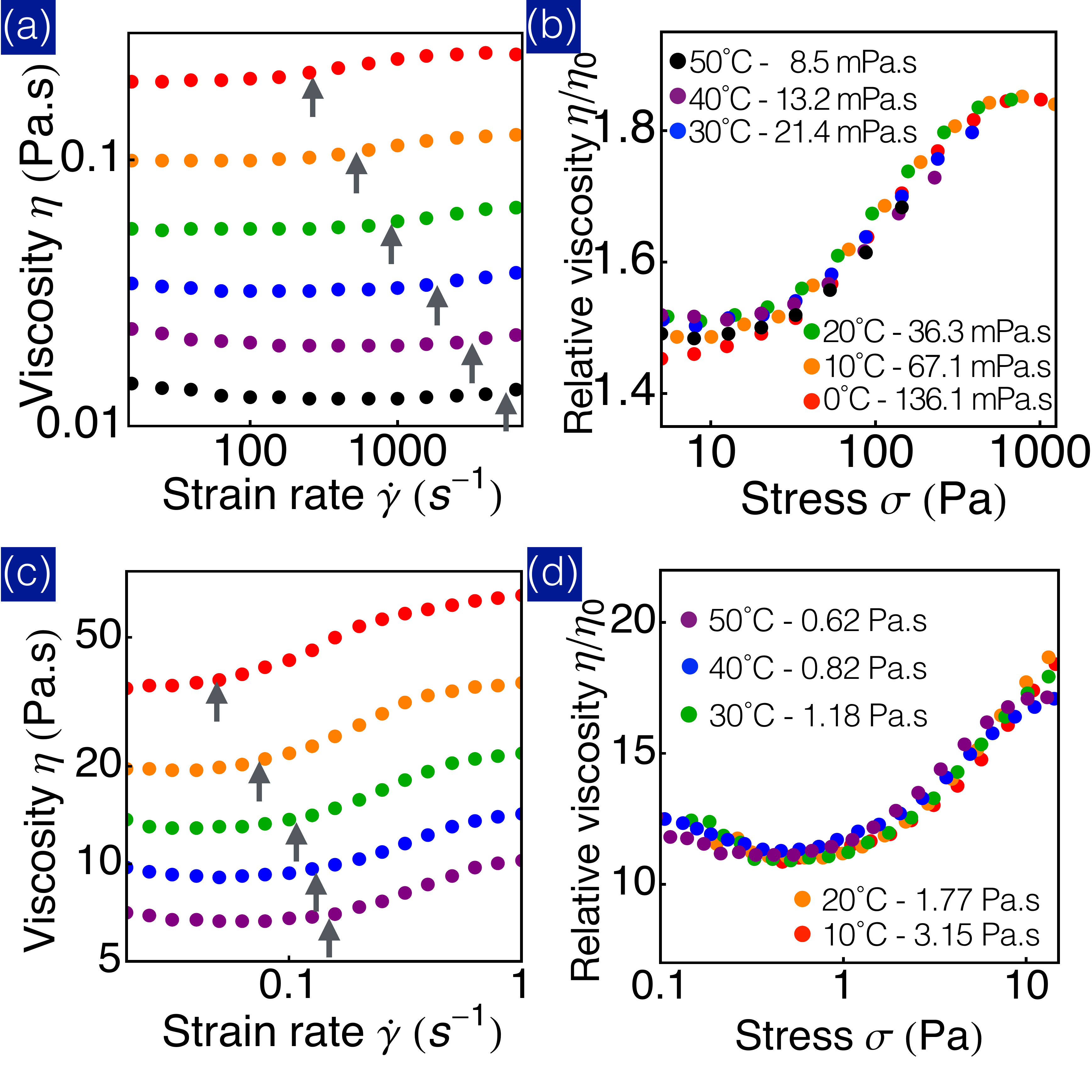}
\caption{Scaling of the relative viscosity with shear stress. (a) and (c): Viscosity
versus shear rate for silica ($\phi=0.43$) and PMMA ($\phi=0.46$) suspensions
at different temperatures. Solvent viscosities and temperatures are as labeled in (b) and (d) (for silica, we used a different solvent composition to figure \ref{fig:fig1_schematics}; see Supplementary Information). Due to the temperature dependence of the solvent viscosity the
initial suspension viscosity and onset strain rate (arrows) of the thickening are
substantially altered. (b,d) Relative viscosity {\it vs} stress showing collapse
of the thickening data onto a master curve. 
\label{fig:temp}
}\end{figure}

We find that as the solvent viscosity varies, the onset of thickening (dashed lines in figure~\ref{fig:temp}(a) and (c)) occurs at different strain rates for both of our suspensions. Thus, in the silica suspension,  figure~\ref{fig:temp}(a), the highest solvent viscosity (lowest temperature) data set shows shear thickening at $\dot\gamma \gtrsim 200$~s$^{-1}$, but this onset progressively moves to higher rates as we lower the solvent viscosity, until in the lowest viscosity data set, we barely see thickening over our range of shear rates, but observe a small degree of shear thinning instead. Similarly, in the PMMA suspension, figure~\ref{fig:temp}(c), we see the onset of thickening at around 0.05~s$^{-1}$ in the highest viscosity data set; this onset also moves to higher shear rates as the solvent viscosity decreases. 

Each of these two data sets, however, can be scaled onto a master curve if we plot the relative viscosity $\eta(\gdot)=\sigma/(\gdot \eta_0(T))$ against the shear stress $\sigma$ \footnote{This is in contrast to confined, sedimenting granular systems under oscillatory shear, where the severity of shear thickening is found to vary with $\eta_0$; Q. Xu, S. Majumdar, E. Brown, and H. M. Jaeger, Euro. Phys. Lett. \textbf{107}, 680004 (2014).}. The master curve for the silica particles, figure~\ref{fig:temp}(b), indeed shows a single onset stress at $\approx 10$~Pa, while that for the PMMA system, figure~\ref{fig:temp}(d), shows an onset stress at $\approx 1$~Pa. Thus, in both systems, our experiments show that shear thickening occurs above a critical stress. The difference in the magnitude of the onset stress of the two suspensions is consistent with the expectation that the load needed to press particles into contact should be sensitive to details of the stabilization mechanism \cite{Fernandez2013}. 

Our observation of contact-driven shear thickening contradicts the conclusions drawn from rheo-SANS \cite{Gurnon2015} and scattering dichroism measurements on Brownian hard spheres \cite{Bender1996}. The rheo-SANS data provides an approximation of the hydrodynamic stresses through a non-trivial analysis of the structural anisotropy in the suspension. However, even the latest rheo-SANS measurements show thickening that is only a small fraction of that observed in rheometry \cite{Gurnon2015}. In the scattering dichroism measurements, the total suspension viscosity, $\eta$, and the viscosity contribution from Brownian stresses, $\eta_B$, can be measured independently. The authors found that the contribution to the viscosity from other stresses, $\eta-\eta_B$, increased upon shear thickening, and attributed this to an increase in hydrodynamic stresses \cite{Bender1996}. However, the remaining viscosity may in fact be dominated by contact contributions, an interpretation that is in alignment with results from dense non-Brownian suspensions \cite{Lemaitre2009} as well as the data presented in the current work. 

Scattering dichroism measurements \cite{D'Haene1993} also found evidence of particle clusters in the shear-thickened state  (later confirmed by real-space imaging \cite{Cheng2011}), which were supposed to originate from large lubrication stresses between particles. The origin of these clusters remains unclear, however -- the present work suggests that any clustering is likely to involve frictional contacts.

To conclude, we have used shear reversal to show that, for two canonical model colloidal systems, continuous shear thickening does not originate from hydrodynamic interactions but from the formation of particle contacts. The onset of contact formation, and thus shear thickening, is found to occur above a critical stress whose value is sensitive to whether the particles are charge- or sterically-stabilized. Further work is needed to establish whether shear thickening of nanoparticles, for which Brownian stresses are not negligible, is also driven by contact formation; this should be possible with shear reversal. 

While shear reversal alone provides no \emph{a priori} information about the nature of the interactions between contacting particles, the quantitative agreement between the experimental and simulation viscosity response after reversal, figure 2(b), (d) and 3(a), strongly suggests that static friction is present. Indeed, evidence for frictional contacts was recently found in a similar model system \cite{Guy2015}.  A careful study of the effect of particle surface properties -- specifically surface roughness -- on suspension rheology would be required to validate this claim, which in turn motivates the need for robust measurements of particle friction.

We thank Gareth McKinley for use of his modified ARES rheometer, John Brady and Mike Cates for fruitful discussions of the shear reversal method, the Cohen group for helpful inputs and Andy Schofield for synthesizing PMMA particles. The Cornell work was supported by NSF CBET-PMP Award No.~1232666 and continued support from NSF CBET-PMP Award
No.~1509308. BMG and CN held EPSRC CASE studentships with Johnson Matthey. MH and WCKP were funded by EPSRC grant EP/J007404/1.


\end{document}